\documentstyle[preprint,aps]{revtex}
\begin{document}
\draft
\preprint{ISU--NP--94--15}
\title{Renormalization of Effective Hamiltonians}
\author{T.J.~Fields, K.S.~Gupta, and J.P.~Vary}
\address{Department of Physics and Astronomy,\\
Iowa State University of Science and Technology\\
Ames, Iowa  50011, USA}
\date{\today}
\maketitle

\begin{abstract}
We introduce a way of implementing Wilson renormalization within the
context of
the theory of effective Hamiltonians.
Our renormalization scheme involves manipulations
at the level of the generalized $G$--matrix and is independent of any
specific kinematics. We show how to calculate the beta function
within this
context and exhibit our method using simple scale--invariant quantum
mechanical
systems.
\end{abstract}

The theory of effective Hamiltonians and operators in many--body
physics has a
long and elaborate history \cite{gmatrix}.
This framework primarily involves manipulation of
operators and is independent of the chosen kinematics.
Our present effort is aimed at
incorporating the concepts of renormalization within the context of
the theory of effective Hamiltonians.
Recently, other approaches for Hamiltonian renormalization have
appeared both in field theory \cite{wil2} and quantum mechanics
\cite{pinsky,wil3}.

Our general philosophy towards
renormalization is inspired by the original work of Wilson
\cite{wilson}.
Wilson's approach involved
integrating out degrees of freedom above a certain momentum range to
arrive at
an effective action. This concept has primarily been implemented in
the
path integral formulation of a given problem. In recent years there
has
been a renewed effort to implement
renormalization within a Hamiltonian formulation
\cite{wil2,pinsky,wil3}, as this may lead to a better understanding
of
key issues in strong interaction physics.

	In this Letter we present a new non--perturbative
scheme for renormalization which utilizes some of the more recent
developments in the theory of effective Hamiltonians for many--body
systems.  Although our approach is quite general, we illustrate the
methods with simple one--body quantum mechanical problems.
In this way, we can exhibit the efficacy of our ideas without being
overwhelmed by technical complications.

Consider the eigenvalue problem
\begin{equation}
H | \Psi_{i} \rangle =E_{i} | \Psi_{i} \rangle
\; \; \; \; i=1,2,\ldots,n,
\label{orig}
\end{equation}
with $n$ possibly infinite.
There are very few Hamiltonians for which Eq. (\ref{orig})
is exactly solvable.
In addition, the usual perturbation theory may turn out to be
inadequate, and
a large number of degrees of freedom associated with a many--body
system may
prevent a straightforward analysis of the problem. In realistic
situations one
would, in general, seek a subset
of all the solutions of the above Hamiltonian. This naturally leads
to the
concept of an effective Hamiltonian.

We shall now briefly describe our construction of the effective
Hamiltonian.
We can arbitrarily split up any Hamiltonian $H$ as
\begin{equation}
H=H_0+V,
\end{equation}
where $H_0$ is exactly solvable.
One motivation for doing
this step is to provide a convenient basis to work with, defined by
the set of eigenvectors of $H_0$:
\begin{equation}
H_0 | \mu \rangle =E_{\mu} | \mu \rangle .
\end{equation}
All matrices will be written with respect to this $| \mu \rangle$
basis,
unless otherwise indicated.

We choose a
model space ${\cal M}$ which contains $d$ basis vectors
of $H_0$.
The operators $P$ and $Q$ which project into
and out of our model space, respectively,  are given by
\begin{eqnarray}
P & = & \sum_{\mu \subset {\cal M}} | \mu \rangle \langle \mu |,
\nonumber \\
Q & = & \sum_{\mu / \! \! \! \!  \subset{\cal M}}
| \mu \rangle \langle \mu | = 1 - P.
\end{eqnarray}
It is helpful to think of these projection operators
as partitioned matrices :
\begin{eqnarray}
P & = &
\left (
\begin{array}{c|c}
1 & 0\\ \hline
0 & 0\\
\end{array}
\right ) , \nonumber \\
Q & = &
\left (
\begin{array}{c|c}
0 & 0\\ \hline
0 & 1\\
\end{array}
\right ).
\end{eqnarray}
Consider, as in \cite{ls}, a transformation of our Hamiltonian
\begin{eqnarray}
\widetilde{H} & = & e^{-S}He^{S} \nonumber, \\
| \widetilde{\Psi}_{i} \rangle & = & e^{-S} | \Psi_i \rangle,
\end{eqnarray}
where $S$ is an operator to be determined shortly.
It follows directly from this transformation that the eigenvalues
of $\widetilde{H}$ are the same as the eigenvalues of the
original Hamiltonian, i.e.,
\begin{equation}
\widetilde{H}| \widetilde{\Psi}_{i} \rangle =E_i
| \widetilde{\Psi}_{i}\rangle.
\end{equation}
We will use the freedom in the choice of $S$ to require that
\begin{equation}
\widetilde{H}(P | \widetilde{\Psi_i}\rangle )=E_i(P
| \widetilde{\Psi_i} \rangle ) \; \; \; \; i=1,2,\ldots,d.
\label{goal}
\end{equation}
As expressed in conventional applications to many--body
problems, the goal of the effective Hamiltonian formalism
is to construct an operator which acts
only in a model space, yet gives us
a subset of the exact eigenvalues of
the full Hamiltonian.  For our purposes here, we restate the
goal as that of obtaining a subset of well--defined solutions
of the eigenvalue problem for $H$.

For this purpose, we choose the effective Hamiltonian as
\begin{equation}
H_{\rm eff} \equiv P\widetilde{H}P,
\label{heff}
\end{equation}
which clearly acts only on the states in the model space
(which may, in itself, be infinite dimensional).
{}From Eqs. (\ref{goal}) and (\ref{heff}) it also follows that
\begin{equation}
H_{\rm eff}(P | \widetilde{\Psi_i} \rangle )=E_i(P
| \widetilde{\Psi_i}\rangle ).
\end{equation}
The expression in Eq. (\ref{heff}) is therefore a consistent choice
for
$H_{\rm eff}$.

The problem of finding $H_{\rm eff}$ now reduces to one of finding an
appropriate $S$. Following \cite{ls,zvb}, we choose to obtain $S$
such that
$S=QSP$.  This implies
that in our chosen basis $S$ must have the form
\begin{equation}
S=
\left (
\begin{array}{c|c}
0 & 0\\ \hline
\hat{s} & 0\\
\end{array}
\right )
\end{equation}
where $\hat{s}$ is a (presently) arbitrary $n-d$ by $d$ matrix.

By choosing $S$ in this particular way, $S^n$ is zero for all
$n > 1$.
This leads to
\begin{equation}
e^{S}=1+S=
\left (
\begin{array}{c|c}
1 & 0\\ \hline
\hat{s} & 1\\
\end{array}
\right )
\end{equation}
with an analogous result for $e^{-S}$.

If we started out with an arbitrary hermitian Hamiltonian
\begin{equation}
H=
\left (
\begin{array}{c|c}
a & b\\ \hline
b^{\dagger} & f\\
\end{array}
\right ).
\end{equation}
with $a=a^{\dagger}$ and $f=f^{\dagger}$, then
\begin{eqnarray}
\widetilde{H} & = & e^{-S}He^{S} \nonumber \\
              & = &
\left (
\begin{array}{c|c}
a+b\hat{s} & b\\ \hline
-\hat{s}(a+b\hat{s})+b^{\dagger}+f\hat{s} & f-\hat{s}b\\
\end{array}
\right )
\end{eqnarray}
The effective Hamiltonian $H_{\rm eff}$ therefore takes the form
\begin{equation}
H_{\rm eff} = P\widetilde{H}P = a+b\hat{s}.
\end{equation}

We shall now exhibit an iterative method to obtain $H_{\rm eff}$.
Following \cite{zvb} we define
\begin{equation}
Z \equiv  H_{\rm eff}-\omega = a+b\hat{s}-\omega,
\label{z}
\end{equation}
which is equal to the effective Hamiltonian up to the
arbitrary additive constant $\omega$.
Next we explicitly split
$H$ into $H_0 + V$ so that
\begin{eqnarray}
H_{0} & = &
\left (
\begin{array}{c|c}
\lambda_{P} & 0\\ \hline
0 & \lambda_{Q}\\
\end{array}
\right ) , \nonumber \\
V     & = &
\left (
\begin{array}{c|c}
a-\lambda_{P} & b\\ \hline
b^{\dagger} & f-\lambda_{Q}\\
\end{array}
\right ),
\label{matr}
\end{eqnarray}
where $\lambda_{P}$ And $\lambda_{Q}$ are the (diagonal) matrices
containing the eigenvalues of $H_0$.
Finally we introduce
a generalized $G$--matrix defined as [6-7]
\begin{eqnarray}
G(\omega) & \equiv & PVP + PVQ \frac{1}{\omega - QHQ}QVP \nonumber \\
& = & PVP+PVQ\frac{1}{\omega-QH_{0}Q}QVP +\nonumber \\
&&+PVQ\frac{1}{\omega-QH_{0}Q}
QVQ\frac{1}{\omega-QH_{0}Q}QVP+ \cdots ,
\label{gexp}
\end{eqnarray}
which, for the above conventions, can be written as
\begin{equation}
G(\omega) = (a-\lambda_{P})+b\frac{1}{\omega - f}b^{\dagger}.
\label{g}
\end{equation}
Eqs. ({\ref{z}) and (\ref{gexp}) can be solved iteratively to give
$Z$ and $G(\omega)$.  From now on, we will denote $G(\omega)$ as $G$.
One such iteration scheme is \cite{ls}
\begin{eqnarray}
Z_1 & = & PH_{0}P + G - \omega P, \nonumber \\
Z_{n} & = &\frac{1}{1-G_1-G_2Z_{n-1}-G_3Z_{n-2}Z_{n-1}-\cdots -
G_{n-1}Z_2Z_3\cdots Z_{n-1}}Z_1,
\end{eqnarray}
where
\begin{equation}
G_k(\omega)=\frac{1}{k!}\frac{d^k}{d\omega^k}G(\omega).
\end{equation}
$H_{\rm eff}$ can finally be constructed from the above solution for
$Z$.

We note in passing that the generalized G--matrix may provide a
leading
approximation to $H_{\rm eff}$.  Within that approximation, our Eq.
(\ref{gexp}) bears resemblance to the effective Hamiltonian
introduced
in \cite{pinsky}.

	We shall now introduce the concept of renormalization within the
above
framework. We have seen above that the knowledge of the matrix $G$
allows us to
obtain $Z$, which is identical to $H_{\rm eff}$ up to an additive
constant. In
what follows, we shall therefore restrict our attention only to $G$.
For the sake of convenience we choose to work in the momentum
representation where the kinetic energy term in the Hamiltonian
is diagonal.
To introduce the concept of renormalization we shall
focus our attention on the one--particle system. The formal
generalization
to a many--particle system would be straightforward.
The matrix elements of $G$ are here given by
\begin{eqnarray}
G_{kk'} &=&  \langle k | PVP | k' \rangle + \nonumber \\
&&+\int \,dp  \,dp' \langle k | PVQ | p \rangle \langle p |
\frac{1}{\omega-QH_0Q}
| p' \rangle \langle p' | QVP | k' \rangle  + \cdots
\label{gmat}
\end{eqnarray}

Let us suppose that the potential $V$ depends on
a single coupling constant $\mu_0$, which we shall call the bare
coupling
constant.
It is clear from Eq. (\ref{gmat})
that the matrix element $G_{kk'}$ will be a function of $\mu_0$.
The expression in Eq. (\ref{gmat}) may, in general, require
regularization
due to the divergence arising from the integral.
The regularization that we choose consists of introducing an
ultraviolet
cutoff $\Lambda$.
The matrix element in Eq. (\ref{gmat}) is now a function
of the coupling constant $\mu_0$ and the cutoff
$\Lambda$. At the end of the calculation we must remove the cutoff,
i.e. we must take $\Lambda$ to $\infty$, which, as discussed above,
may in
general lead to divergence. One way to avoid the divergence is to
replace
the coupling constant $\mu_0$ with a function of $\Lambda$, which we
denote as $\mu (\Lambda)$,
and then require that
matrix element in Eq. (\ref{gmat}) remain finite and independent
of the cutoff as the cutoff is removed.
In other words, we demand that
\begin{equation}
\lim_{\Lambda \to \infty}\frac{d}{d \Lambda}G_{kk'}(\Lambda, \mu
(\Lambda))=0.
\label{der}
\end{equation}
The function $\mu (\Lambda)$ thus plays the role of the renormalized
coupling
constant.

The dependence of the coupling constant on the cutoff is usually
expressed in terms of the beta function, which is defined by
\begin{equation}
\beta (\mu) \equiv \Lambda \frac{d \mu}{d \Lambda}.
\label{beta}
\end{equation}
Within our formalism, Eqs. (\ref{der}) and (\ref{beta}) can be used
to
calculate the beta function.

Note that once Eq. (\ref{der}) is satisfied and $\mu(\Lambda)$ is
determined, then $H_{\rm eff}$ (via $Z$), based on $G_{kk'}(\Lambda,
\mu(\Lambda))$, should also be independent of $\Lambda$ as
$\Lambda \to \infty$.  Thus, the complete problem of
renormalization is solved.

We shall now illustrate the method prescribed above in two simple
cases of a Dirac particle in 1 dimension and a Schrodinger particle
in 2
dimensions \cite{pinsky,huang}.
In both these cases the
interaction potential will be taken as a delta function in position
space :
\begin{equation}
V(x) = -\mu_0\delta^{(n)}(x),
\end{equation}
where $n$ is the dimension of configuration space.
In the momentum space the interaction potential would simply be a
constant, i.e.,
\begin{equation}
V(k) = - \mu_0.
\end{equation}
We will choose $H_0$ to be the pure kinetic operator, and our model
space to consist of all states with
momenta less than $\lambda$.  Thus $Q$ projects onto the momentum
range $[\lambda,\infty]$.

	With the choice of the interaction potential described above, the
series in Eq. (\ref{gmat}) can be summed exactly and is given by
\begin{equation}
G_{kk'}=\frac{-\mu_{0}}{1+\mu_{0}I(\omega)}\delta(k-k'),
\end{equation}
where $I(\omega)$ is given by
\begin{equation}
I(\omega)\equiv \int_{\lambda}^{\infty}\,d^{n}p\frac{1}
{\omega-E_{0}(p)}.
\label{I}
\end{equation}
Following the preceeding discussion we now introduce an ultraviolet
cutoff
$\Lambda$.
Replacing $\mu_0$
by the renormalized coupling constant $\mu$ and using Eqs.
(\ref{der}) and
(\ref{beta}),
we obtain the beta function as
\begin{equation}
\beta(\mu) = \mu^2 \Lambda \frac{\partial I}{\partial \Lambda}.
\end{equation}

To obtain the explicit expression for the beta function we need to
evaluate the integral appearing in Eq. (\ref{I}). For the 1
dimensional Dirac
particle we have $n=1$, $E_{0}(p)=p+m$ and
\begin{equation}
I(\omega)\equiv \int_{\lambda}^{\Lambda}\,dp\frac{1}{\omega-(p+m)}
=-\,{\rm
ln}\left(\frac{\omega-(\Lambda+m)}{\omega-(\lambda+m)}\right).
\end{equation}
The corresponding beta function is given by
\begin{equation}
\beta (\mu) = -\mu^2.
\end{equation}

For the Schrodinger particle in 2 dimensions we have
$n=2$ and $E_{0}(p)=p^2/2$ (we set the mass of the particle as 1.)
Proceeding exactly as before, we obtain
\begin{equation}
I(\omega)=-2\pi  \,{\rm ln}\left(\frac{\omega-\Lambda^2}
{\omega-\lambda^2}\right)
\end{equation}
and
\begin{equation}
\beta = -4\pi \mu^2.
\end{equation}

Note that the results in both examples above have the desirable
property that the beta function is independent of the model space
cutoff, $\lambda$. The beta functions calculated give rise to
asymptotically free theories and generate the generally accepted
pattern for the flow of the coupling constant for the two examples
described above.

In conclusion, we have introduced a way of implementing a
nonperturbative renormalization scheme within the context of
many--body effective Hamiltonian theory.
We have tested the method with applications to simple
scale--invariant quantum mechanical systems.
These examples exhibit the efficacy of our
ideas and calculations for more realistic many--body systems and for
quantum field theory are presently under
investigation.

\acknowledgements

We acknowledge helpful discussions with S.S.~Pinsky.
This work was supported in part by the U.S. Department of Energy
under Grant No. DEFG02-87ER40371, Division of High Energy and Nuclear
Physics.  J.P.V. acknowledges support from the Alexander von Humboldt
Foundation.

\end{document}